\documentclass[english]{lni}

\IfFileExists{latin1.sty}{\usepackage{latin1}}{\usepackage{isolatin1}}

\usepackage{graphicx}
\usepackage{fancyhdr}
\usepackage{listings} 
\usepackage{changepage} 
\usepackage[figurename=Fig., tablename=Tab., small]{caption}[2008/04/01]

\author{Asim Abdulkhaleq\footnote{Institute of Software Technology, University of Stuttgart, Asim.Abdulkhaleq@informatik.uni-stuttgart.de}, ~Stefan Wagner\footnote{Institute of Software Technology, University of Stuttgart,  Stefan.Wagner@informatik.uni-stuttgart.de}, ~Daniel Lammering \footnote{Continental, Regensburg, Germany, Daniel.Lammering@continental-corporation.com}, ~Hagen Boehmert\footnote{Continental, Frankfurt am Main, Germany, Hagen.Boehmert@continental-corporation.com } ~ and Pierre Blueher \footnote{Continental, Frankfurt am Main, Germany, Pierre.Blueher@continental-corporation.com} }
\title{Using STPA in Compliance with ISO 26262 for Developing a Safe Architecture for Fully Automated Vehicles }
\begin{document}
\maketitle
\renewcommand{\refname}{References}
\setcounter{footnote}{2} 

\begin{abstract}
 Safety has become of paramount importance 
 in the development lifecycle of the modern automobile systems. However, the current automotive safety standard ISO 26262 does not specify clearly the methods for safety analysis. Different methods are recommended for this purpose. FTA (Fault Tree Analysis) and FMEA (Failure Mode and Effects Analysis) are used in the most recent ISO 26262 applications to identify component failures, errors and faults that lead to specific hazards (in the presence of faults). However, these methods are based on reliability theory, and they are not adequate to address new hazards caused by dysfunctional component interactions, software failure or human error. A holistic approach was developed called STPA (Systems-Theoretic Process Analysis) which addresses more types of hazards and treats safety as a dynamic control problem rather than an individual component failure. STPA also addresses types of hazardous causes in the absence of failure. Accordingly, there is a need for investigating hazard analysis techniques like STPA. In this paper, we present a concept on how to use STPA to extend the safety scope of ISO 26262 and support the Hazard Analysis and Risk Assessments (HARA) process. We applied the proposed concept to a current project of a fully automated vehicle at Continental. As a result, we identified 24 system-level accidents, 176 hazards,  27 unsafe control actions, and 129 unsafe scenarios. We conclude that STPA is an effective and efficient approach to derive detailed safety constraints. STPA can support the functional safety engineers to evaluate the architectural design of fully automated vehicles and build the functional safety concept.

\end{abstract}

\section{Introduction}
  Nowadays, innovations in software and technology lead
  to increasingly complex automotive systems such as self-parking vehicles,  the use of smartphones to park vehicles and, more recently, fully automated driving vehicles. As a new technology, the fully automated driving vehicles may bring a new safety risk and threats to our society which have to be controlled during their development.   
  Hence, the safety analysis becomes a great challenge in the development of safety-critical systems. In the past, failures of the automotive systems beyond separate component malfunction like interface problems led to safety issues. The automotive industry started to pay attention to the functional safety of vehicle electronic control systems and to introduce new standards to address the growing complexity of its systems. The safety standard ISO 26262 \cite{ISO2011} ``Road vehicles -- Functional safety" is an international risk based safety standard to describe state-of-the-art for the development of safety-relevant
  vehicle functions and addresses possible hazards caused by malfunctioning behaviour of electrical/electronic systems.
   
\textbf{1.1 Problem Statement:}
The ISO 26262 is a document intended to achieve functional safety regarding an E/E component. It specifically addresses hazards resulting from the presence of failures and malfunctions emanating from hardware (HW) and software (SW). These may be introduced by random (HW-related) failures/malfunctions and/or systematic SW failures. The intention is to give arguments for functional safety if ``best practice" guidelines have been followed during concept and development phases. The Hazard Analysis and Risk Assessment (HARA) \cite{ISO2011}  defines possible hazards; deductive and inductive analyses look for E/E faults and failures that lead to these hazards. In contrast to the general assumption, following this guideline does not mean that the product is ``safe". It is the authors' belief that a ``safe"  product does not only result from the absence of hazards during the presence of malfunctions, but also the absence of hazards in the absence of malfunctions. That bears the question \textbf{how these hazards can be identified and what their cause might be}.

\textbf{1.2 Research Objectives:}
This research work answers that question by using the STPA method as an approach to identify the potential hazards of fully automated vehicles in compliance  with ISO 26262 at an early concept phase and provide  safety constraints on how the risk for an accident can be mitigated by avoiding those hazards.  

\textbf{1.3 Contribution:}
We provide guidance on how to use STPA in compliance with ISO 26262. We apply STPA to the existing architecture design of a fully automated driving vehicle to develop a safety concept  to enhance the architecture design.

\textbf{1.4 Context:}
This work was conducted in the form of a cooperation between Continental, which is a German automotive manufacturing company, and  the University of Stuttgart, during the development process of a fully automated driving vehicle project.

\textbf{1.5 Terminology:}

\textbf{Functional Safety} is ``Absence of unreasonable risk due to hazards caused by malfunctioning behavior of Electrical/Electronic systems"\cite{ISO2011}.

\textbf{Operational Safety (Roadworthiness):} is ``a property or ability of any kind of automobile to be in a suitable operating condition or meeting acceptable standards for safe driving and transport of people, baggage or cargo in roads or streets" \cite{opertionalsafety2014}.

\textbf{Item:} ``is a system or array of systems or a function to which ISO 26262 is applied" \cite{ISO2011}.

\section{Background}
\subsection{Hazard Analysis Approach: STPA}
STPA (Systems-Theoretic Processes Analysis) \cite{leveson2011engineering}
was developed by Leveson in 2004 based on the STAMP  (Systems-Theoretic Accident Model and Processes) causality accident model  for identifying system hazards and safety-related constraints necessary to ensure acceptable risk in complex systems.  STPA helps to identify causal factors and unsafe scenarios in which the safety constraints can be violated. STPA results in identifying a larger set of causes, many of them not involving failures or unreliability, while traditional techniques were designed to prevent hazards due to component failure accidents (caused by one or more components that fail). The main steps of STPA are divided into three sub-steps:  (1) Establish the fundamentals of
the analysis (e.g. system-level accidents and the associated
hazards) and draw the control structure diagram of the
system. (2) Use the control structure
diagram to identify the potentially unsafe control actions. (3)
Determine how each potentially unsafe control action (accident causes) could
occur by identifying the process model and its variables
for each controller and analysing each path in the control
structure diagram. 

STPA has been successfully applied and extended in different domains such as STPA for automotive systems \cite{Abdulkhaleq2013}, STPA for cybersecurity \cite{Young:2014:IAS:2556647.2556938} and STPA for software safety \cite{Abdulkhaleq20152}.

\subsection{ISO26262 Safety Standard }
ISO 26262 (Road vehicles functional safety) \cite{ISO2011} is an international functional safety standard, which  provides guidance, recommendation and argumentation for a safety-driven product development in the automotive area. Safety classification and suggestions for specific safety development processes may aid to stipulate functional safety for each new product as State-of-the-Art.

\begin{figure*}[!t]
	\centering
	\includegraphics[width=\textwidth ]{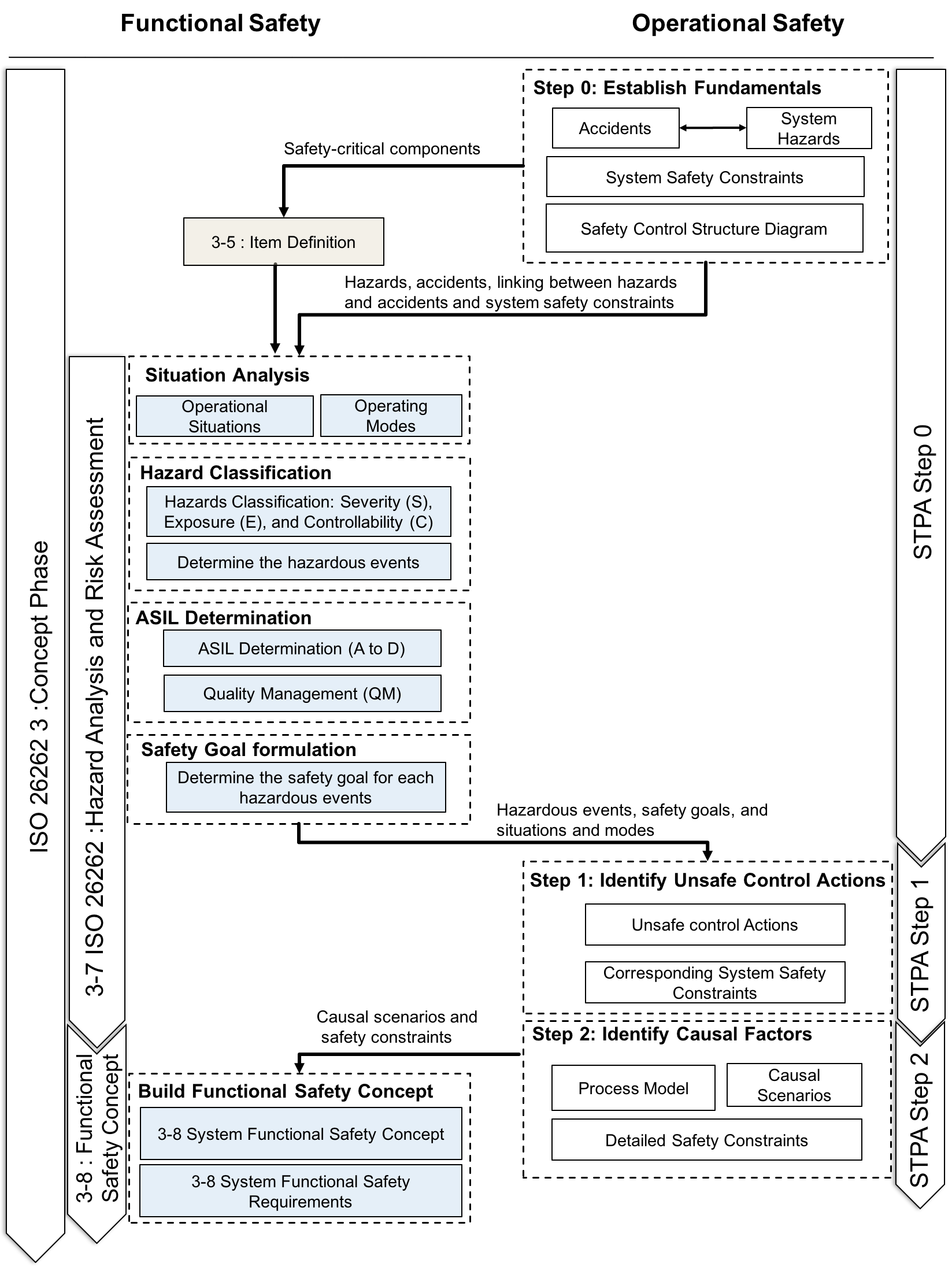}
	\caption{ Integration of STPA into the ISO26262- part 3 concept phase}
	\label{fig_sim}
\end{figure*}

ISO 26262 is structured into 10 parts and describes the safety activities in 7 parts (3--9).
 Part 3 specifics the concept phase (as shown in Fig. \ref{fig_sim}) which starts with defining the item (e.g. system, array or function) and performing the hazard and risk analysis for the item. The results are the safety goals for all hazards which are derived and classified with an ASIL (Automotive Safety Integrity Level) rating which is a risk classification scheme defined by the ISO26262 standard. Part 4 specifies the requirements for product development at the system level for automotive applications. Parts 5 and 6 specify the requirements for product development at the hardware level and software level for automotive applications. Part 7 specifies the requirements for production, operation, service and decommissioning.Part 8 specifies the supporting processes (e.g. hardware and software tool qualification). Part 9 specifies the automotive safety integrity level (ASIL)-oriented and safety-oriented analysis.

\section{Related Work}

Hommes \cite{hommes2012review, hommes2015} highlighted the benefits of applying STPA in the automotive domain and using STPA as a hazard analysis technique in the ISO 26262 lifecycle, especially in the concept phase (ISO26262 part 3). Hommes also provided an assessment review of the ISO 26262 standard's ability to address the challenges in ensuring the safety of complex software intensive E/E systems. Hommes mentioned that there is a lack in guidance on hazard identification and elimination in the concept phase, which makes the ISO 26262 standard not sufficient to provide safety assurance. That gives us a strong motivation to investigate the use of STPA in compliance with the ISO 26262 standard to gain a deeper understanding about the benefits and limitations of using STPA as a hazard analysis approach to support the HARA process in ISO 26262.

Recently, Mallya et al. \cite{Mallya2016} analysed how STPA can be used in an ISO 26262 compliant process. They mapped every relevant activity and artifact required or recommended by the HARA process which can be satisfied by applying STPA. They emphasized that the key difference between STPA and HARA is the risk assessment process.  However, both STPA and HARA have different base assumptions for identifying hazards. Based on this work, We explored how STPA can extend the safety scope of ISO 26262 by considering different factors that cause inadequate controls during the operational time of a vehicle (e.g. human, environment). We also proposed a concept on how to use STPA to support HARA activities in ISO 26262 instead of mapping STPA activities onto the HARA process.   

We have applied STPA to a well-known example of a safety-critical systems in the automotive domain \cite{Abdulkhaleq2013}: the Adaptive Cruise Control system (ACC). This case study was performed based on an existing case study with MAN Truck \& Bus AG \cite{5635049} in which the authors conducted an exploratory case study applying safety cases for the ACC system. We compared the results of STAMP/STPA with the safety cases on the same system. In \cite{abdulkhaleq2014software}, we proposed a safety verification methodology based on STPA safety analysis. We applied STPA to vehicle cruise control software to identify the software safety requirements at the system level and verify these safety requirements at the design level.  Recently, we proposed a safety engineering approach for software-intensive systems based on STPA \cite{Abdulkhaleq20152}, called \emph{STPA SwISs} which combines the activities of safety engineering and software engineering. Thus, in turn, it shall help to reduce the associated software risks to a low level. This approach can be applied in compliance with the ISO 26262 part 6 at the software level.

\begin{figure}[t]
	\centering
	\includegraphics[width=3.6in]{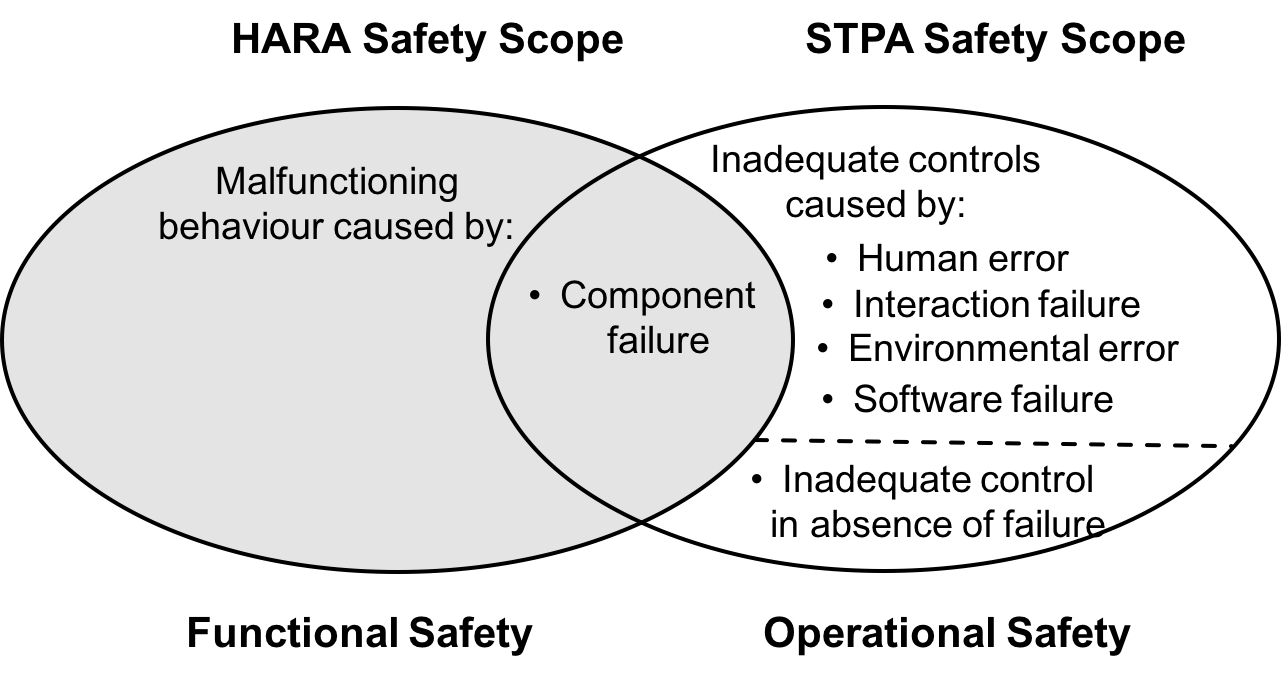}
	\caption{The safety scope of STPA and  HARA in ISO 26262}
	\label{fig_Scope}
\end{figure}

\begin{table*} [p]
	\centering
	\def\arraystretch{1.0}
	\caption{The terms used in STPA and the terms used in ISO 26262 part 3: concept phase }
	\label{tabterm}
	\begin{tabular}{ p{6.0cm}     p{6.0cm}   }
		
		STPA Terminologies \cite{   leveson2011engineering}                                                                                                                                & ISO  26262   Terminologies  \cite{ISO2011}
		\\ \hline
		
		\textbf{Accident}:  Accident (Loss) results from inadequate enforcement of the behavioural safety constraints on the process.
		& \textbf{ No corresponding  term}      \\   
		\textbf{No corresponding term }        & \textbf{Harm:}  is a  physical injury or damage to the health of persons.
		\\   
		
		\textbf{Hazard} is a system state or set of conditions that, together with a particular set of worst case
		environmental conditions, will lead to an accident.
		&      \textbf{Hazard } is  a potential source of harm caused by malfunctioning behaviour of the item.

		\\    	
		\textbf{No Corresponding} & \textbf{Item} is a system or array of systems to implement a function at the vehicle level, to which ISO 26262 is applied.
		\\ 
		\textbf{Unsafe Control Actions} are the hazardous scenarios which might occur in the system due to a provided or not provided control action when it was required.   & \textbf{ No corresponding term}  
		\\
		\textbf{ No corresponding term}  & \textbf{Malfunctioning Behaviour}: is a failure (termination of the ability of an element to perform a function as required) or unintended behavior of an item with respect to its design intent
		\\  
		
		\textbf{Safety Constraints} are the safeguards which prevent the system from leading to losses (accidents)  &     	\textbf{Functional Safety Requirements}: are specifications of implementation-independent safety behaviour, or implementation-independent safety
		measures, including its safety-related attributes.
		
		\\ 		 
		\textbf{Causal Factors}  are  the accident scenarios that explain how unsafe control actions might occur and how safe control actions might not be followed or executed.  &   \textbf{ No corresponding term}\\

		\textbf{ No corresponding term} &   \textbf{Hazardous Events} are combinations of a hazard and an operational situation.
		
		\\  
		\textbf{Corresponding safety Constraints} are top-level safety constraints which are derived based on the unsafe control actions& \textbf{Safety Goals} are top-level safety requirements as a results of the hazard analysis and risk assessments  \\
		
		\textbf{No Corresponding term}  & \textbf{Functional Safety Concept} consists of functional requirements and preliminary architectural assumptions. \\
		
		\textbf{Process Model} is a model required to determine the environmental \& system variables (process model variables) that affect the safety of the control actions.  & (Partially) \textbf{Operation Situation} is a scenario that can occur during a vehicle's life. \textbf{Operating Mode} is
		perceivable functional state of an item or element.	 \textbf{Safe State} is operating mode of an item without an unreasonable level of risk (e.g. switched-off mode, intended operating mode, degraded operating mode)			  \\

		\textbf{No Corresponding term}    & \textbf{ASIL} is one of four levels to specify the item's or element's necessary requirements of ISO 26262 and safety measures to apply for avoiding an unreasonable residual risk, with D representing the most stringent and A the least stringent level 				  \\
		
		\hline
		
	\end{tabular}
	
\end{table*}

\section{The Concept of Using  STPA in the ISO 26262 Lifecycle}

The main goal of STPA is to identify inadequate control scenarios which can lead to accidents and develop detailed safety constraints to eliminate and control these unsafe scenarios. The main starting point of STPA is to identify potential accidents and hazards at the system level and draw hierarchical safety control structure of the system. The main activities of the concept phase in ISO 26262 are defining an item, identifying the hazardous events that need to be eliminated or controlled and developing the safety concept at the system level.  
The Hazard Analysis and Risk Assessment (HARA) process in ISO 26262 consists of the following activities \cite{ISO2011}: (1) Situation Analysis and Hazard identification, (2) Hazard classification, (3) Hazard determination, and (4) Safety goal determination.  In this paper, we used STPA as a hazard analysis technique to support HARA by defining an item and identifying the hazards and unsafe scenarios of the item at the system level. The term ''item" is used here to refer to the system.

Figure \ref{fig_Scope} shows the safety scope of both STPA and HARA. STPA focuses on identifying the potential inadequate controls that could lead to the hazards. The inadequate control can be caused by human error, interaction failure, environmental, software failure. STPA also focuses on identifying the inadequate controls in absence of the individual component failures (e.g. dysfunctional interactions or unhandled conditions). The safety scope of the HARA in ISO 26262 is to identify the possible hazards caused by the malfunctioning behaviour of electronic and electrical systems (individual components). To use STPA in the ISO 26262 lifecycle, we first define the important terms of STPA and ISO 26262 (shown in Table \ref{tabterm}). Based on our expertise in STPA and ISO 26262, we map the terms of STPA and ISO 26262 which have the same meaning.  In the following, we summarise the main steps to use STPA in compliance with ISO 26262 in  the part 3 concept phase (shown in Fig. \ref{fig_sim}):
\begin{enumerate}
	\item [1.]Apply STPA Step 0 (Fundamentals Analysis):
	\begin{enumerate}
		\item [1.1.] Identify Accidents and system-level hazards.
		\item [1.2.]Identify the high-level system safety constraints.
		\item [1.3.] Draw the control structure diagram of the system 
	\end{enumerate}
	\item [2.] Use the results of  STPA Step 0 to define an item and item information needed  (e.g. purpose, content of item, functional requirements etc.). The control structure diagram in STPA Step 0 shows the main components which form a system under analysis. This diagram contains information to help the functional safety engineer to define an item and its boundaries. 
	\item [3. ] Use the list of hazards, accident, the high-level system safety constraints identified in STPA Step 0 as an input to the HARA approach. 
	\item [4 ] Apply the HARA approach:
	\begin{enumerate}
		\item [4.1] Determine the operational situations and operating modes in which an item's malfunctioning behaviour may lead to potential hazards. 
		\item [4.2] Classify the hazards identified in Step 0 based on the estimation of three factors: Severity (S), Probability of Exposure (E) and Controllability (C)
		\item Identity the hazardous events by considering the hazards in different situations. 
		\item [4.3] Determine ASIL (Automotive Safety Integrity Levels) for each hazardous event by using four ASILs: A (the lowest safety integrity level) to D (the highest one). If the hazardous event is not unreasonable, we refer it as QM (Quality Management).
		\item [4.4] Formulate the safety goal for each hazardous event.   
	\end{enumerate}
	\item [5.]  Use the hazardous events, safety goals, situations and modes as input to the STPA Step 1. 
	\item [6.] Apply STPA Step 1 to identify the unsafe control actions of an item 
	\item [7.] Apply STPA Step 2 to identify the causal factors and unsafe scenarios of each unsafe control action identified in STPA Step 1.
	\item [8.] Use the results of STPA Step 1 \& 2  to develop the  system functional safety concept and safety requirements at this level.
	
	

	
\end{enumerate}



\section{Application Example: Fully Automated Driving Vehicle}

\begin{figure}[!t]
	\centering
	\includegraphics[width=\textwidth,  height=3.0 in]{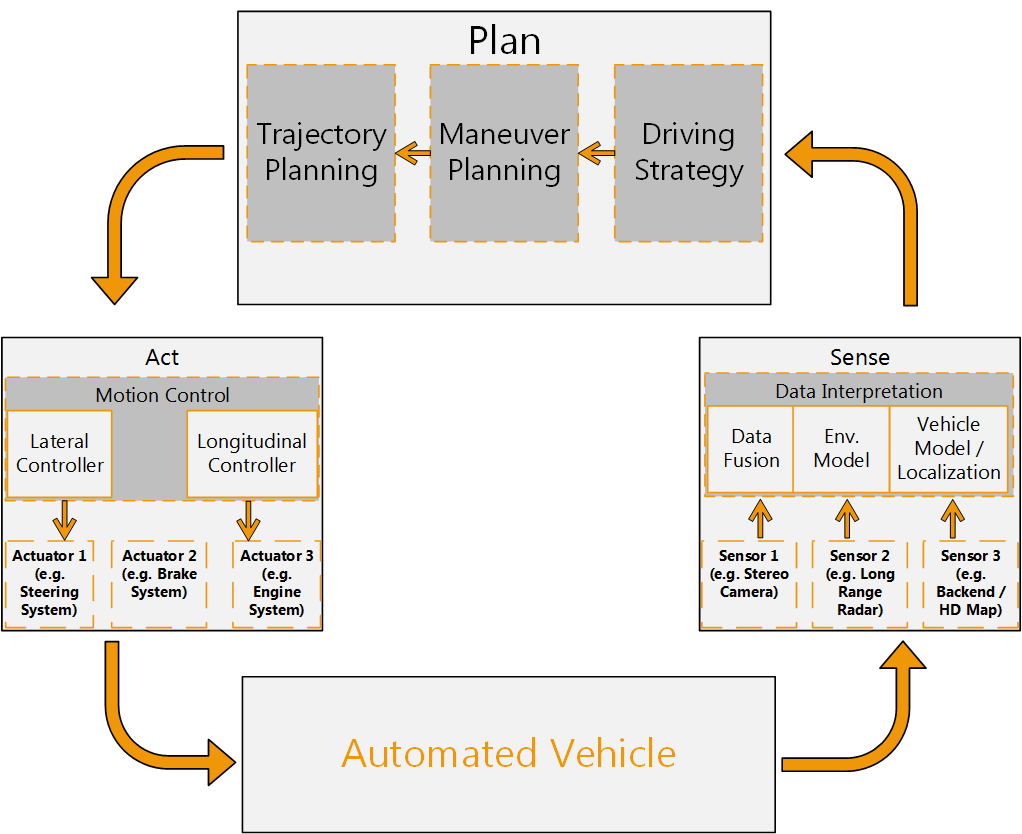}
	\caption{ Functional Architecture of Fully Automated Driving Vehicles}
	\label{fig_sim2}
\end{figure}

\textbf{Study Object:}
Semi-automatic and fully automated driving requires compliance with essential system features like reliability, availability, security and safety. A fully automated driving system (SAE Level 5) \cite{SAE2016} involves the act of navigating the car without any input from the human driver through the use of sensing the environment, performing and calculating a desired driving path (trajectory) and sending the desired controls to the actuators (as shown in Fig. \ref{fig_sim2}). Therefore, the required components for a fully automated driving system can be classified in three main groups: 1) \textbf{Sense}: Several sensors are necessary to gather information of the environment, perception of the vehicle state and traffic participants. For example, wheel speed sensors, cameras, short and long range radars and even lidar technology are used to get a fine-grained environment model; 2) \textbf {Plan:} The planning component consist of multiple levels of decision making. The driving strategy plans upcoming maneuvers and is a core element of the car's calculated behaviour. It serves as an input for the trajectory planning sub-module. The function of the trajectory planning is processing a safe vehicle trajectory with respect to the surroundings (collision avoidance) and to the given maneuver. The motion controller calculates the motion requests from the desired trajectory. 3) \textbf {Act:} In the Act group all needed actuators, e.g. brake system, steering system and engine control, for longitudinal and lateral movement can be summarized. The output of the motion controller provides the torque requests for the actuators.

\begin{table}[]
	\centering
	\def\arraystretch{1.3}

	\caption{Examples of the system level accidents }
	\label{acc}
	\begin{tabular}{p{1cm} p{10cm}}
		
		ID        & Accident                                                                                                        \\ \hline
		1         & AD vehicle lost the steering and collided into an object moving in front on a highway.                                    \\
		2         & AD vehicle lost the steering/braking suddenly while the vehicle moving up in the hill and made an accident. \\
		3         & AD vehicle made a collision due to loss the communication signals with Backend.                             \\
		4         & AD vehicle collided into an object or vehicle due to a wrong driving strategy                               \\
	  \hline                                     
	\end{tabular}
\end{table}

\textbf{Apply the STPA Step 0: Fundamentals Analysis}
First, we do the first part of STPA Step 0 (identifying the system-level accidents). As a result, we identified 24 system-level accidents which the fully automated driving system can lead or contribute to. Table \ref{acc} shows examples of the system level accidents. For example, \emph{AC.1: The fully automated vehicle collided into an object moving in front on a highway.}  Second, we do part 2 of Step 0 (identifying hazards). As a result, we identified 176 hazards which can lead to these accidents. For example, a hazard can lead to accident \emph{AC.1}, \emph{HA.1: The fully automated vehicle lost steering control because it received wrong ego longitudinal torque.} Third, we do part 3 of Step 0 (identifying the system-level safety constraints).  An example for a high-level system safety constraint is \emph{SC.1: The fully automated driving vehicle must receive correct data all the time while driving on a road.}

\begin{table}[]
\centering
\def\arraystretch{1.3}

\caption{Examples of the system level hazards}
	\label{haz}
	\begin{tabular}{p{1cm} p{10cm}}
		ID & Hazards                                                                   \\  \hline
		1  & The AD vehicle lost steering control because it received wrong ego longitudinal torque. \\
		2  & The AD vehicle does not detect a moving obstacles in the front.           \\
		3  &   The AD vehicle moves with no data of prediction of situation and scenario for traffic participants. \\
		4  & The AD vehicle receives wrong environmental model data.  \\  \hline                
	\end{tabular}
\end{table}

\begin{figure}[h]
	\centering
	\includegraphics[width=\textwidth, height=3.0 in]{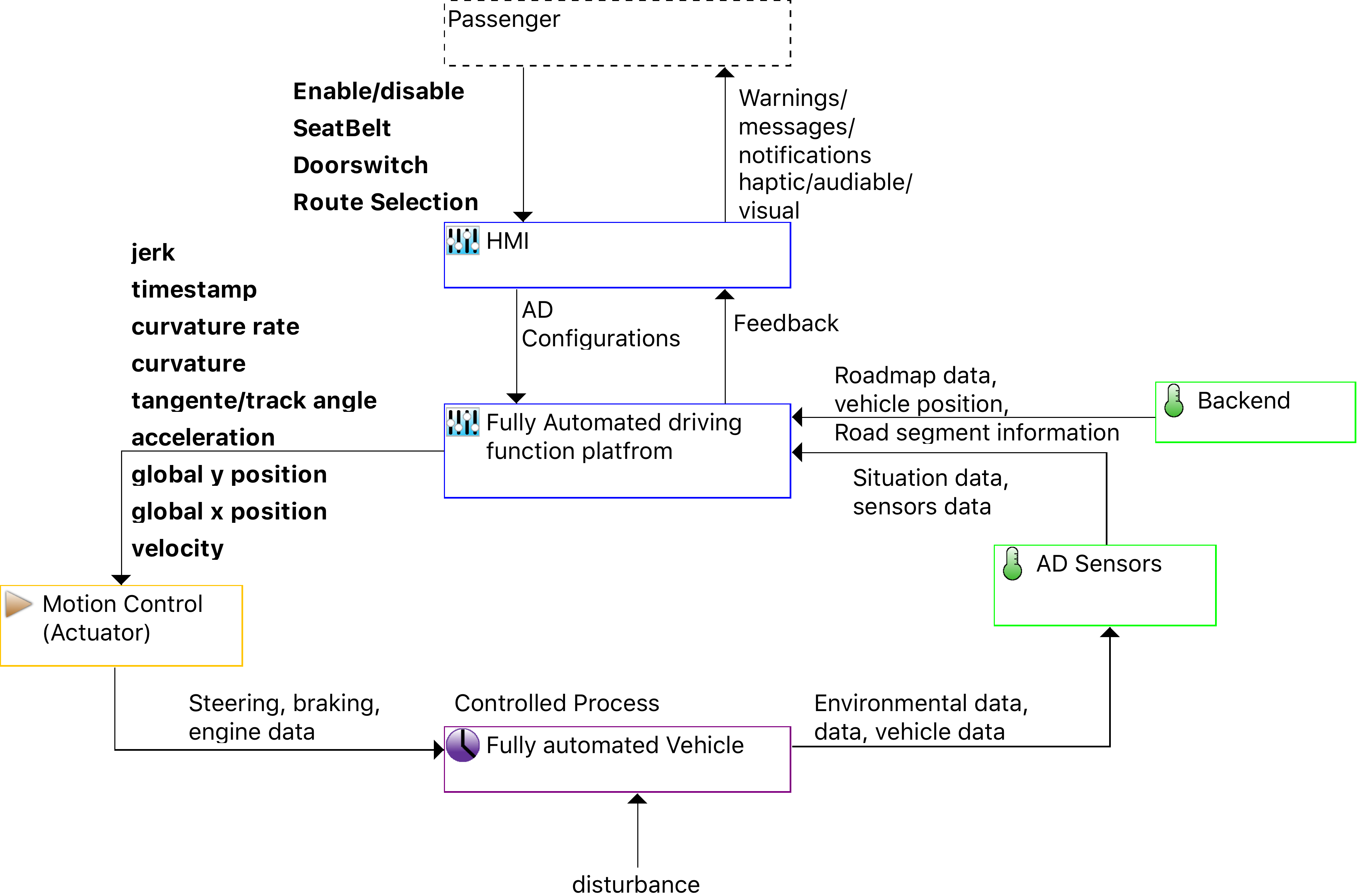}
	\caption{The safety control structure diagram of fully automated driving system }
	\label{fig_scd}
\end{figure}

Fourth, we do the part 4 of Step 0 (drawing  the high-level safety control structure diagram). Figure \ref{fig_scd}) shows the control structure diagram of the fully automated driving system at the architectural design level. The control structure diagram shows the main components which interact with the fully automated driving system in the vehicle. We used the results of the STPA Step 0 to define the item. For example, the fully automated driving function platform is an item in which the ISO 26262 can be applied. The control structure diagram shows the boundary of the fully automated driving function platform and its interfaces. The purpose of the fully automated driving function platform is to control the autonomous vehicle by issuing the control actions to the motion control and receiving the feedback from the different sensors.  

\textbf{Using the STPA Step 0 results:} 
We also used the results of the STPA Step 0 (list of accidents, list of hazards) as an input to the HARA approach. First, we determine the operational situations and operating modes. For example, the operation situation from the accident AC.1 can be determined as \emph{OS1: crashing on a highway}. The operating mode is \emph{OM.1: driving}. We classified each hazard identified in the STPA Step 0 with two factors (severity and exposure). For example, we estimated the severity for the hazard HA1 as S3 (Life-threatening injuries or fatal injuries), the probability of exposure of the operational situation (on highway) as E3 (Medium probability). Then, we identified an hazardous event HE.1 from the hazard HA.1 and the accident AC.1 as \emph{HE.1: The fully automated vehicle lost control of steering while driving on a highway.} Next, we estimated the controllability for each hazardous event (C0: simply controllable to C3: difficult to control). The controllability \cite{monkhouse2015notion} is a way of assessing the likelihood that the hazardous situation is usually controllable or not. As assumptions of the fully automated vehicles (SAE level 5), the driver is not expected to take control at any time. Therefore, we assigned the controllability as C3 (Difficult to control or uncontrollable) for each hazardous events of fully automated driving system.  For example, the controllability of the hazardous event HE.1 is C3. We also assigned an ASIL to each hazardous event. For example, the ASIL of the hazardous event HE.1 is ASIL C. We formulated the safety goal for each hazardous events. For example,  the safety goal for the hazardous event HE.1. is  \emph{SG.1: the fully automated driving vehicle must not lose the steering control while driving on a highway}.

\textbf{Apply the STPA Steps 1 \& 2:} 
We used the output of the HARA approach (e.g. list of hazardous events and operational situations) as an input to the STPA Step 1 to formulate the unsafe control actions. We used the control structure diagram to identify the unsafe control actions of the fully automated driving vehicle (STPA Step 1). To identify unsafe control actions, we first identified the critical safety control actions at a high level of abstraction. For example, the fully automated driving function platform has a control action called \emph{trajectory}.  Primarily, the trajectory contains a time sequence of state-space points with \emph {timestamp, x and y position, velocity, acceleration, track angle, jerk, curvature and curvature rate}. The trajectory is issued by the automated driving function platform to the motion controller.

We evaluated each of these control actions within four general hazardous types \cite{leveson2011engineering} (e.g. not providing, providing incorrect, providing at wrong timing/order, and stopped too soon/applied too long) to check whether or not they lead to hazardous events. We identified 27 unsafe control actions. For example, \emph{UCA-1: The fully automated driving function platform does not provide a valid trajectory to motion control while driving too fast on a highway}. This unsafe control action (malfunctioning behaviour) can lead to the potential hazard HA.1.   

To generate the corresponding safety constraints, we translated each unsafe control action into a corresponding safety constraint by using the guide words e.g. ``shall" or ``must".  For example, a corresponding safety constraint  SC-1 for unsafe control action UCA-1 is: \emph{ The fully automated function platform must always provide a valid trajectory to motion control while driving on a highway}.
We used the results of the situation analysis to determine the process model of the Automated Driving (AD) function platform. For example, a process model variable of the AD function platform is the \emph{road\_type} which has the following values: \emph{highway, parking, intersection, mountain, city, urban}.  We used the results of STPA Steps 0 \& 1 as input to STPA Step 2 to identify the causal factors and scenarios. We also determined the accident causes (STPA Step 2) for each unsafe control action to get a deeper understanding on how they could occur in the fully automated driving vehicle.  For example, a causal scenario for the unsafe control action UCA-1 is: \emph{ The fully automated driving function platform receives wrong signals from backend due to the lack of communication while driving too fast on a highway}. Then, a new safety constraint can be derived as \emph{The fully automated driving function platform shall receive correct data from the backend without delay during driving.} 

\textbf{Using the STPA Step 2 results:} 
We used the results of the STPA Step 2 to build the safety concept and addressed the new safety requirements. For example, the causal scenario CS.1 of UCA-1 is: \emph{The AD function platform does not provide a valid trajectory to motion control while the system is active and the vehicle is moving and there is traffic ahead on highway}. Then, we identified the safety constraints (SC) for each causal scenario. For example, the safety constraint (SC.1) for CS.1 is: \emph{the AD function platform must always provide the trajectory to enable motion control to adjust the throttle position and apply brake friction when the vehicle is moving and there is traffic ahead to avoid a potential collision}. We used the results of the STPA Step 2 to build the functional safety concept and determine the functional safety requirements.

\section{Discussion}
Based on our work, we found that STPA and HARA have different base assumptions. HARA has two parts: 1) Hazard analysis which aims at identifying the hazards that lead to harm. However, these hazards are related to the individual component failures and they are not described in terms of other accident causes such as interaction failures between vehicle and its environment and driver (a passenger in a fully automated vehicle). The second part is the risk assessment which aims at identifying risks of each identified hazard. Whereas STPA focuses on control problems, not component failures. STPA aims at identifying inadequate control caused by component failures, human errors, and component interaction errors among the system components. It is also able to identify hazards that arise due to unsafe interactions among the system components in the absence of  component failures. Therefore, STPA can identify more types of hazards and not only hazards which may occur due to the component failures, but also the hazards which may occur in the absence of component failures.  To fill this gap between HARA and STPA, we showed how to use the results of STPA as in input to support the HARA process activities instead of mapping the HARA activities and artifact to STPA. Moreover, STPA does not support risk analysis while HARA supports this kind of activities. The work here shows that STPA and HARA are complementary to each other and STPA can be used to extend the safety scope of ISO 26262.

Another gap in the concept phase in ISO 26262 is that there is no systematic way to define the item \cite{kannan2015analysis}. We found that STPA can fill this gap by applying the STPA Step 0 before starting the concept phase. The STPA Step 0 can define the item and establish the information needed for the item. Moreover, the STPA Step 0 can define the safety constraints for each item. 

The STPA Step 2 requires that defining the process model of the controller, which determines the current state of the controlled process and how the control actions will be issued by the controller. This model also is used to determine the causal scenarios of each unsafe control action identified in STPA Step 1.  However, there is no guidance on how to define the process model and its variables which should be augmented in the process model. We figured out that HARA can fill this gap by using the situation analysis which determines the operational situations and operating modes in which an item's malfunctioning behaviour will result in a hazardous event. These situations and modes can be used as input to the process model into STPA.

An assumption of STPA is that it can be applied at all stages of system development process, especially at an early stage \cite{leveson2011engineering}. This assumption is similar to the assumption of the HARA process which can also be applied at an early stage of system development. However, STPA helps to derive more detailed safety requirements, not only the functional safety requirements which are the main output of the concept phase in ISO 26262. Therefore, mapping the results of STPA Step 1 and Step 2 to build the functional safety concept requires high expertise in both STPA and HARA. Furthermore, STPA is a top-down process and the detailed design of item is not necessary to be known before applying STPA to the item. This assumption is also similar to the assumption of the HARA process  which can be applied with a little bit of knowledge about the detailed design of the item.

In conclusion, we believe that STPA can support the HARA process in ISO 26262 activities and help the functional safety engineers to develop the functional safety requirements for each item identified at an early stage in the concept phase based on the results of the STPA Step 0. Indeed, the integration of STPA into HARA activities still needs modification in the assumptions and terms of both STPA and HARA to directly map the results of STPA into HARA.
\section{Conclusion}
In this paper, we explored the use of the STPA approach as hazard analysis in compliance with ISO 26262 to improve the safety architecture of the fully automated driving vehicle project at Continental. Our work showed that STPA is a powerful hazard analysis technique which can be used to support the safety lifecycle and HARA process in ISO 26262 by providing a systematic guidance on defining an item, deriving detailed safety constraints and developing safety goals and a functional safety concept. That helps us to evaluate the architectural design of the new fully automated driving system at an early stage of the development process.  As future work, we plan to explore the use of the STPA approach in compliance with ISO 26262 at different levels of the fully automated driving architecture (e.g. software level) to develop detailed safety requirements. We plan also to conduct an empirical case study evaluating our proposed concept with functional safety engineers at Continental to understand the benefits and limitations of using STPA to support the HARA process and to extend the safety scope of ISO26262. We plan also to develop an extension to XSTAMPP \cite{abdulkhaleq2016xstampp2} to support the HARA process activities.

\bibliographystyle{lnig}
\bibliography{Literatur.bib}

\end{document}